# Study of the Electron Mobility in InAs/GaSb Type II Superlattices at High Temperatures

**S. Safa[1][a], A Asgari [a,b]**

[a]Research Institute for Applied Physics and Astronomy, University of Tabriz,

Tabriz 51665-163, Iran

[b]School of Electrical, Electronic and Computer Engineering, The University of Western Australia,

Crawley, WA 6009, Australia

**Abstract:**

In this paper, we present a study of the effects of temperature on the electron mobility in InAs/GaSb type-II superlattices (SLs) in which the band structures and wave functions are calculated by solving the K.P Hamiltonian using the numerical Finite Difference method. In the model the dominant scattering mechanisms such as alloy scattering, acoustic phonon scattering and optical phonon scattering are taken into account. The obtained electron mobility of the type II SLs is depended on the structural parameters and different scattering parameters. A comparison of our calculated results with published experimental data is shown to be in good agreement.

## 1. Introduction

Type II InAs/GaSb SLs was proposed by Sai-Halasz and L. Esaki in the 1970's [1]. The cutoff wavelength of these SLs can be tailored from 3µm to 30µm, by varying the thickness of the InAs or GaSb layers, covering mid, long, and very long wavelength regions. In the past few years, type-II InAs/GaSb SLs have been developed quickly, especially because of their important role in opto-electronic devices [2-8]. However their transport properties still have a big lack with respect to the theory. The theory of diffusive carrier transport in SLs was developed by Mori and Ando [9], Dharssi and Butcher [10], and others [11, 12, 13, 14]. Although, the effects of phonon scattering on electronic transport in semiconductors and type-I SLs have been generally studied [15, 16, 17, 18], no investigation of such effects has been carried out before on the particular case of type-II SLs by considering the relation to structural parameters, alloy concentration and temperature.

---

[1] Author to whom correspondence should be addressed. Electronic mail:
safa@tabrizu.ac.ir. Tel.: 98-41-33393010. Fax: 98-41-33347050.

There are two important effects that have not been studied comprehensively in type II SLs yet: first, alloys created in the SL interfaces, limit the electron mobility and as the second, phonon scattering leads to a significant decrease of the electric conductivity at high temperatures. The purpose of this paper is to report on results of electron mobility calculations in type II InAs/GaSb SLs by considering the scattering mechanisms that limit mobility in these systems at high temperatures. For this purpose, first, the band structures and wave functions have been calculated by the use of the numerical Finite Difference method. This technique is based on constructing the K.P Hamiltonian matrix of the finite system from the properties of each individual layer in the structure. In these calculations, the precise value of each SL layer has been included in periodic boundary conditions, so the exact wave functions for all interfaces are determined. In the next step, the electron mobility has been calculated, which is based on the solution of Boltzmann's equation. Finally, the effects of temperature and SL structural parameters on the mobility have been analyzed. The comparison of calculated electron mobility and band structure parameters with the experimental data shows that our theoretical results are in good agreement with the experimental measurements.

**2. Analytical Formalism**

*2.1. Electronic structure*

As a numerical technique, a straightforward theoretical finite difference method has been developed and subsequently applied to calculate the electronic band structure of InAs/GaSb type-II SLs. The basic principle behind this method is the use of a matrix to express the K.P Hamiltonian, which constitutes any arbitrary number of permitted band orbitals and can also describe, as accurately as possible, the band structure and wave functions of the SL.

$$H\psi_{n,k} = E_{n,k}\psi_{n,k} \qquad (1)$$

where

$$H = \frac{p^2}{2m_0} + \frac{\hbar}{m_0}K.P + \frac{\hbar^2 k^2}{2m_0} + V + \frac{\hbar}{4m_0^2 c^2}(\sigma \times \nabla V)\cdot(\hbar K + P) \qquad (2)$$

, and we name $H_0 \equiv \dfrac{p^2}{2m_0}+V$ in which $V$ is the SL profile potential.

Due to the narrow band gap of type-II materials, it is necessary to use an 8-band coupled Hamiltonian to include the strong interaction among the 8 bands nearest to the band gap, namely the conduction band (CB), the heavy-hole (HH), the light-hole (LH), and the spin-orbit split-off (SO) bands, each with a double spin degeneracy. The basis functions used here are $u_{n,0}\uparrow$ and $u_{n,0}\downarrow$, $n \in \{s, p_x, p_y, p_z\}$, which are the eigenfunctions of Hamiltonian $H_0$. Any other wave function of the electron is a linear expansion in this basis as

$$\psi_{n,k} = \sum_m C_{n,m}\uparrow u_{m,0}\uparrow + \sum_m C_{n,m}\downarrow u_{m,0}\downarrow \qquad (3)$$

Note that the full Hamiltonian will generate the matrix elements $H_{mn} = \langle u_{m,0}|H|u_{n,0}\rangle$ in which $k_z$ is replaced by differential representation $-i\dfrac{\partial}{\partial z}$ and the Finite Difference method is applied to express the first and second order derivatives in a discrete manner [19,20].

### 2.2. Scattering mechanisms

In this paper, we include scattering potential in our model and demonstrate the significance of the scattering mechanisms. Fermi's golden rule is often used to obtain the scattering rates via scattering potential and electron wave functions. The Boltzmann equation must also be solved to obtain the relaxation time of the electrons in a scattering mechanism. Acoustic and optical phonon scatterings are inelastic mechanisms which change the energy of the electron while the alloy scattering is an elastic process. Thus a brief discussion follows to recognize the behavior of each process.

#### 2.2.1. Acoustic phonon Scattering

The deformation potential theory has been proposed by Bardeen and Shockley [21], which is based on the energy change of an electron due to lattice deformation. The energy change is related to the volume change of a crystal, and the electron-phonon interaction Hamiltonian is defined by [15]

$$H_{Ac} = D_{Ac}\frac{\delta V}{V} = D_{Ac}\nabla \cdot u(r)\quad(4)$$

, where $D_{Ac}$ is the deformation potential for electron scattering by acoustic phonons [22].

The displacement vector of an acoustic phonon $\mathbf{u(r)}$ is given by

$$\mathbf{u(r)} = \sum_q \sqrt{\frac{\hbar}{2MN\omega_q}}\mathbf{e_q}\left[a_q e^{i\mathbf{q}\cdot\mathbf{r}} + a^\dagger_q e^{-i\mathbf{q}\cdot\mathbf{r}}\right].\quad(5)$$

Here, $\mathbf{e_q}$ is the unit vector along the displacement direction and $\mathbf{k}-\mathbf{k'}\pm\mathbf{q}=\mathbf{G}$ where $\mathbf{G}$ is the reciprocal lattice vector, the plus sign indicates the absorption of phonon and the minus sign indicates the emission of phonon.

According to Fermi's golden rule, the scattering rate of acoustic phonon is given by [23]

$$P(\mathbf{k},\mathbf{k'}) = \frac{2\pi}{\hbar}|M(\mathbf{k},\mathbf{k'})|^2 \delta(E(\mathbf{k}) - E(\mathbf{k'}) \pm \hbar\omega_q),\quad(6)$$

where

$$|M(\mathbf{k},\mathbf{k'})|^2 = |\langle \mathbf{k}, n_q | H_{Ac} | \mathbf{k'}, n_{q'}\rangle|^2.\quad(7)$$

Thus, the relaxation time is obtained from the scattering rate associated with a SL as

$$\frac{1}{\tau(\mathbf{k})} = \frac{V}{(2\pi)^3}\int(1-\cos\theta)P(\mathbf{k},\mathbf{k'},\theta)d\mathbf{k'}d\theta\quad(8)$$

The matrix element for this interaction Hamiltonian is given by [16]

$$\langle \mathbf{k'}, n_{q'} | H_{Ac} | \mathbf{k}, n_q\rangle = D_{Ac}\sum_\mathbf{q}\sqrt{\frac{\hbar}{2MN\omega_q}}(i\mathbf{e_q}\cdot\mathbf{q})\sqrt{n_q + \frac{1}{2}\mp\frac{1}{2}}I(\mathbf{k},\mathbf{k'})\delta_{\mathbf{k'},\mathbf{k}\pm\mathbf{q}}\quad(9)$$

,where $I(\mathbf{k},\mathbf{k'}) = \Omega\int d^3r\, u^*_{k'}(\mathbf{r})u_k(\mathbf{r})$ and $MN = L^3\rho$ with $\rho$ being the crystal density.

In the case of electron-acoustic phonon scattering, the associated acoustic phonon energy is quite small and we can consider the semi elastic condition. When the condition $\hbar q / K_B T \ll 1$ is fulfilled, one can approximate

$$n_q \ll n_q + 1 \ll \frac{K_B T}{\hbar \omega_q} = \frac{K_B T}{\hbar v_s q} \tag{10}$$

, where $v_s$ is the sound velocity in InAs and GaSb material. Then the scattering rate is given by [17]:

$$P_{Ac}(\mathbf{k},\mathbf{k}') = \frac{2\pi}{\hbar} \int \frac{D_{Ac}^2 K_B T}{2L^3 \rho v_s^2} q^2 \sin\beta \, d\beta \, d\varphi \, dq \times \delta(E(\mathbf{k}') - E(\mathbf{k}) \pm \hbar\omega_q). \tag{11}$$

*2.2.2. Polar optical phonon Scattering*

In the central $\Gamma$ conduction band valley, the non-polar phonon interaction with electron is generally weak, and the polar interaction can be dominant. In other words, the interaction between electrons and non-polar optical phonons do not exist in the non-degenerate bands with extreme at the $\Gamma$ point in the Brillouin zone. The polarization of the dipole field, that accompanies the vibration of the polar mode is given essentially by the effective charge times the displacement. The latter is just the phonon mode amplitude $u_q$. Hence, we can write the polarization as [16]

$$\mathbf{P_q} = \sqrt{\frac{\hbar}{2\gamma V \omega_0}} \mathbf{e_q} \left[ a_q e^{i\mathbf{q}\cdot\mathbf{r}} + a^\dagger_q e^{-i\mathbf{q}\cdot\mathbf{r}} \right] \tag{12}$$

, where $\mathbf{e_q}$ is the polarization unit vector for the mode vibration, $a^\dagger$ and $a$ are creation and annihilation operators of mode q, $\omega_0$ is the optical phonon frequency in SL and the effective interaction parameter is

$$\frac{1}{\gamma} = \omega_0^2 \left( \frac{1}{\varepsilon_\infty} - \frac{1}{\varepsilon_{(0)}} \right) \tag{13}$$

Here, $\varepsilon_\infty$ and $\varepsilon_{(0)}$ are the high- frequency and low- frequency dielectric permittivity.

Then we can obtain the perturbing energy as [24]

$$\delta E = \left(\frac{\hbar e^2}{2\gamma V \omega_0}\right)^{\frac{1}{2}} \frac{q}{q^2 + q_D^2} \left[a_q^\dagger e^{-i\mathbf{q}\cdot\mathbf{r}} - a_q e^{i\mathbf{q}\cdot\mathbf{r}}\right] e^{-i\omega t} \tag{14}$$

In keeping with the use of a simple screening, the harmonic motion of the phonon can lead to the reduction of screening, so that $q_D$ would be smaller than the Debye screening length. The use of Fermi golden rule leads to the matrix element

$$|M(\mathbf{k},\mathbf{q})|^2 = \left(\frac{\hbar e^2}{2\gamma V \omega_0}\right) \frac{q^2}{(q^2 + q_D^2)^2} \left[(N_q + 1)\delta(E_\mathbf{k} - E_\mathbf{k-q} - \hbar\omega_0) + (N_q)\delta(E_\mathbf{k} - E_\mathbf{k-q} + \hbar\omega_0)\right], \tag{15}$$

and the scattering rate can obtain as [17, 18]

$$\mathbf{P}_{Op}(\mathbf{k}) = \left(\frac{m^* e^2}{4\pi \hbar^2 k \gamma \omega_0}\right) \left[(N_q + 1)\int_{q_{-(emition)}}^{q_{+(emition)}} \frac{q^3 \, dq}{(q^2 + q_D^2)^2} + (N_q)\int_{q_{-(absorbtion)}}^{q_{+(absorbtion)}} \frac{q^3 \, dq}{(q^2 + q_D^2)^2}\right] \tag{16}$$

, where $N_q$ is the Bose-Einstein distribution function.

Letting the screening wave number $q_D = 0$, the relaxation time for polar optical phonon scattering is then given by

$$\frac{1}{\tau_{Op}(E)} = \left(\frac{e^2 \omega_0}{4\sqrt{2}\pi\hbar}\right)\left(\frac{1}{\varepsilon_\infty} - \frac{1}{\varepsilon_{(0)}}\right)\frac{\sqrt{m^*}}{\sqrt{E}}\left[(N_q + 1)\left\{\sqrt{1 - \frac{\hbar\omega_0}{E}} + \frac{\hbar\omega_0}{E}\sinh^{-1}\left(\frac{E}{\hbar\omega_0} - 1\right)^{1/2}\right\} + (N_q)\left\{\sqrt{1 + \frac{\hbar\omega_0}{E}} - \frac{\hbar\omega_0}{E}\sinh^{-1}\left(\frac{E}{\hbar\omega_0}\right)^{1/2}\right\}\right]. \tag{17}$$

*2.2.3. Alloy Scattering*

In a compound semiconductor consisting of three or more elements, each of the three elements is expected not to be periodic in the crystal. This non-uniformity results in a local variation of the periodic potential and thus in electron scattering due to the non-uniform potential. We define an ideal type-II SL to be $(\mathbf{AsIn})_m - \mathbf{AsGa}_x\mathbf{In}_{(1-x)} - (\mathbf{SbGa})_n$ with a perfect crystal structure, where each symbol represents a layer of the same kind of atoms and $AsGa_x In_{(1-x)}$ is an alloy in which $x$ is the concentration of **GaAs**.

The alloy potential is Fourier transformed as [26, 27]

$$V_{alloy}(\mathbf{r}) = \sum_q V_{alloy}(\mathbf{q}) \exp(i\mathbf{q}.\mathbf{r}) \qquad (18)$$

The Fourier coefficient $V_{alloy}(\mathbf{q})$ is taken to be the root-mean square of the shift from the average energy and assumed to be independent of $\mathbf{q}$. Defining the Fourier coefficient for **InAs** and **GaSb** layers (the A and B layers) by $V_a$ and $V_b$, the average potential of the Fourier coefficient is given by

$$V_0 = V_a\, x + V_b(1-x). \qquad (19)$$

When the occupation of layer A is changed from $x$ to $x'$, the change in the potential is given by

$$V' - V_0 = (V_a - V_b)(x' - x). \qquad (20)$$

Therefore, the root mean square value of the potential difference is given by

$$H_{All} = |\langle V' - V_0 \rangle| = |V_a - V_b| \left[ \frac{x(1-x)}{N_c} \right]^{\frac{1}{2}} \qquad (21)$$

, where $N_c$ is the number of (As) cations and corresponds to the unit cell number.

The matrix element for the scattering is then written as

$$\langle \mathbf{k}' | H_{All} | \mathbf{k} \rangle = |V_a - V_b| \delta_{\mathbf{k} \pm \mathbf{q}, \mathbf{k}'} \int \left[ \frac{x(z)(1-x(z))}{N_c} \right]^{\frac{1}{2}} |\varphi_\mathbf{k}(z)|^2 |\varphi_{\mathbf{k}'}(z)|^2 dz \qquad (22)$$

and the relaxation time due to alloy scattering can be obtained by the use of equations (7), (8) and (22).

The $|V_a - V_b|$ depends on many factors such as the symmetry of the conduction band and the electron affinity. In this article this parameter has been used as a fitting parameter [28, 29, 30].

After all these discussions, one can achieve the final result, including the effects of all scattering mechanisms, by the use of Matheson's rule as:

$$\frac{1}{\tau_{Total}} = \frac{1}{\tau_{Ac}} + \frac{1}{\tau_{Op}} + \frac{1}{\tau_{All}}. \tag{23}$$

Thus, in order to calculate the total mobility for different SLs, we have

$$\mu = -e \frac{\int \tau_{Total} \left(\frac{\partial E}{\partial k}\right)^2 \left(\frac{\partial f_0}{\partial E}\right)^2 dk}{\int f_0(k) \, dk}. \tag{24}$$

### 3. Results and Discussion

To model the transport properties of the electrons in the InAs/GaSb SLs, we have computed the band structure for 7 different structures with layer widths $l\,(nm)\,\text{InAs}/l'(nm)\,\text{GaSb}$ that $l, l' = 2, 4, 6, 8$ in a way that all infrared spans are included. Knowing the energies and wave functions for each structure, we have calculated the in-plane scattering rates of electrons via the dominant scattering mechanisms at high temperatures.

As an example, the band structure for a typical strained InAs/GaSb SL ($4\,nm$ /$2\,nm$ thicknesses, respectively) has been calculated numerically based on the finite difference method outlined in Section 1, and the results are depicted in Fig. 1. The calculated band gap is 0.260 $eV$ and as is evident from Fig. 1, the heavy-hole band (HH1) is almost flat, indicating that the corresponding hole mass is 0.77 $m_0$. Thus, heavy holes in z direction are essentially immobile, and their transport contribution can be neglected. Also the calculations show that the semiconductor-semimetal transition will eventually take place if the width of the SL layers is continually increasing and the wave function leakage and effective masses become smaller when the barrier layers become thick.

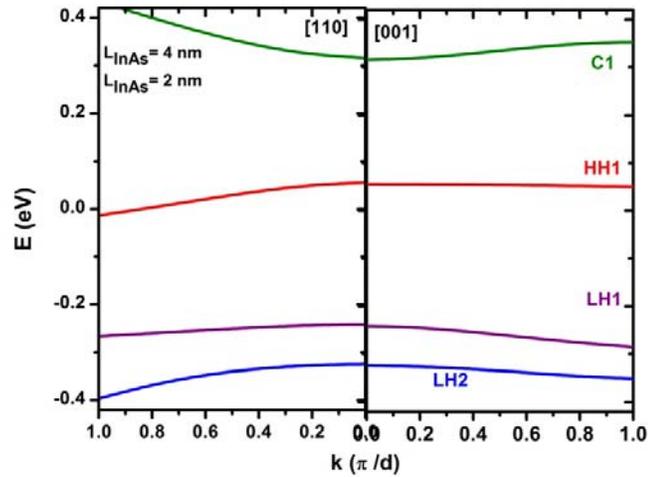

Fig.1.Calculated band structure for (4*nm*/2*nm*) InAs/GaSb SL in (001) and (110) directions.

The samples that have been considered in this discussion consist of 300 periods of InAs and GaSb SLs with different structural parameters.

The electron mobility due to the acoustic phonon scattering has been calculated as a function of temperature for InAs/GaSb SLs with different InAs thickness (Fig. 2. a). As the figure shows and it is well known the acoustic phonon scattering becomes significant at high temperatures which reduces the electron mobility. The electron mobility rises as InAs thickness increase, which can be interpreted by the fact that in SLs with thick layers, the electrons are localized and acquire lower effective mass, which can make the electron mobility relatively high.

The scattering of electrons by optical phonons is one of the main factors determining electron transport features in SLs systems, because at high temperatures the mobility of a carrier is mostly limited by optical phonon scattering. The electron mobility due to the optical phonons scattering for InAs/GaSb SLs has been calculated as a function of temperature for different layer thicknesses, and the results are presented in Fig. 2. b. As depicted in the figure, by increasing the temperature, the mobility decreases rapidly. Also for thick layer SLs, because of lower electron energy (comparable to optical phonon energy), the scattering rates are high and the electron mobilities become severely restricted by optical phonons.

The comparison of two diagrams a and b shows that in thin layer superlattices (2 nm InAs/2 nm GaSb), the acoustic phonon scattering is much effective than optical phonons. This result was expected, because the lower sound velocity in these systems leads to a larger scattering rate. On the other hand, larger effective mass and wave function leakage in thin layer SLs, enhances the acoustic phonon scattering (see Eq. (11)).

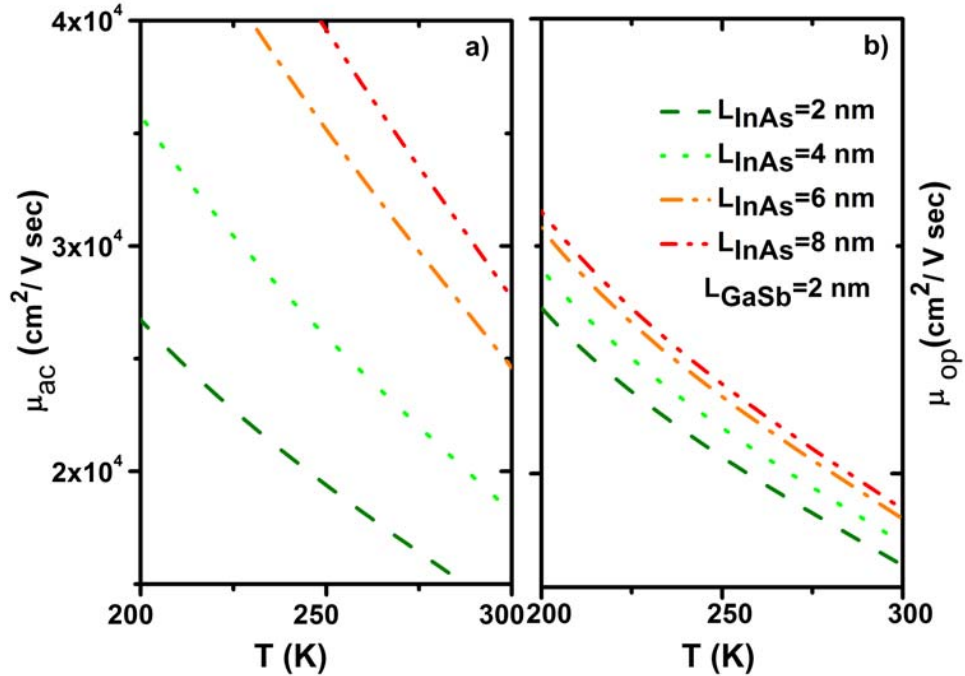

Fig.2. Electron mobility due to a) Acoustic phonon scattering and b) Optical phonon scattering, vs. temperature in InAs/GaSb SLs with different InAs thicknesses.

The electron mobility limited by alloy scattering has been calculated as a function of temperature for different InAs/GaSb SLs structures with various InAs thicknesses and are shown in Fig. 3. As depicted in the figure, this mobility is temperature independent at high temperatures. On the other hand, for structures with thick InAs layers, the mobility is relatively high because of the low effective mass and a small penetration of electrons. Also in thick layer SLs, the effect of alloy concentration is low, according to the equation (22) and the matrix elements become smaller which results in smaller scattering rates.

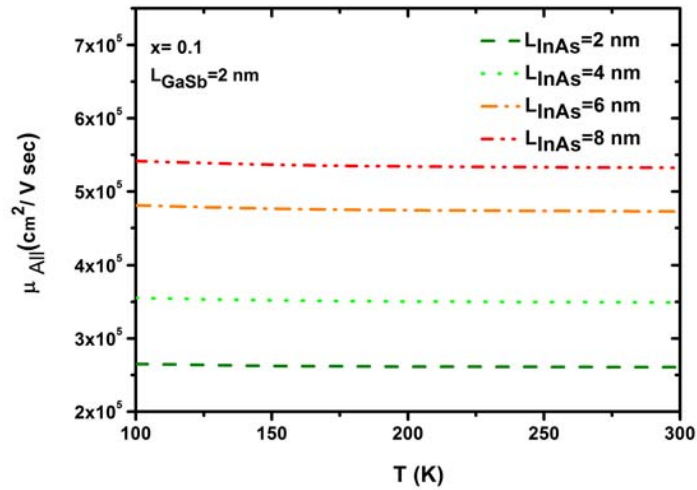

Fig.3. Electron mobility due to alloy phonon scattering vs. temperature in InAs/GaSb SLs with different InAs thicknesses.

The total electron mobility limited by all scattering mechanisms has been calculated as a function of temperature for different InAs/GaSb SLs structures with various InAs thicknesses and are shown in Fig. 4. The behavior of total mobility shows that the electron mobilities drop rapidly with increasing the temperature, which is consistent with $T^{-3/2}$ behavior of bulk materials. It has been shown that the mobility increases generally by increasing the thickness of wells. This figure shows that the electron mobility is mostly limited by phonon scattering mechanisms at high temperatures.

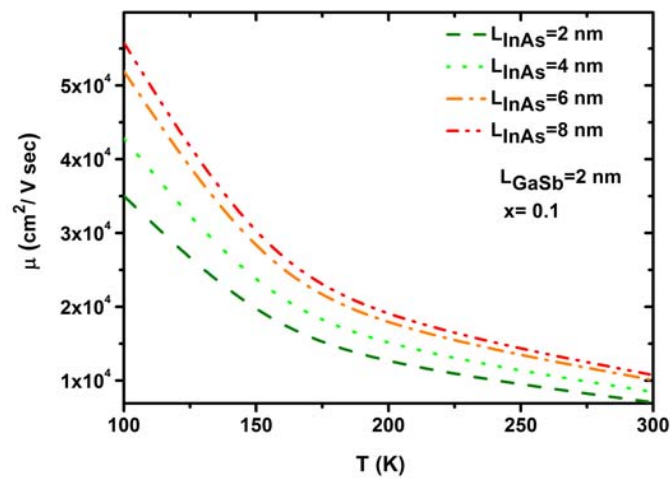

Fig.4. Total electron mobility vs. temperature in InAs/GaSb SLs with different InAs thicknesses.

To confirm the accuracy of our model, we examined it for an InAs/GaSb SLs which has been studied experimentally in Ref [31]. Both InAs and GaSb layers in this SL have 2.7 *nm* thicknesses. We have solved the K.P Hamiltonian for this SL through the Finite Difference method in order to find the wave functions and band structure. After then, the formalisms mentioned in section 2 are applied to this structure. The solid red line in Fig. 5. shows the summation of three scattering mechanisms with Matheson's rule. In order to fit the theoretical results to the experimental data, the alloy concentration considered in calculations must be about *x=0.1*.

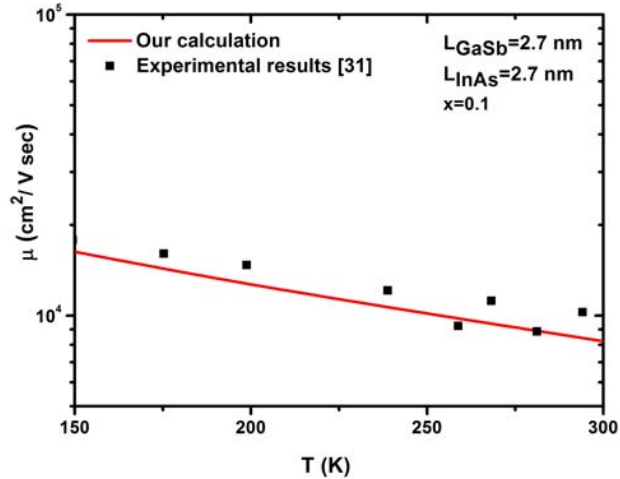

Fig.5. Total electron mobility vs. temperature in InAs/GaSb SLs with 2.7 *nm* InAs, and 2.7 *nm* GaSb. Experimental data are taken from Ref [31].

Also, we compared our results with other experimental data in Ref. [32] which is for a InAs/GaSb SL structure with a thickness of 2.1 *nm* for GaSb layers and a thickness of 3.9 *nm* for InAs ones, and a growth temperature about $435^0$C.

Fig. 6 shows the results of three scattering mechanisms and also the total mobility which is in good agreements with experimental results.

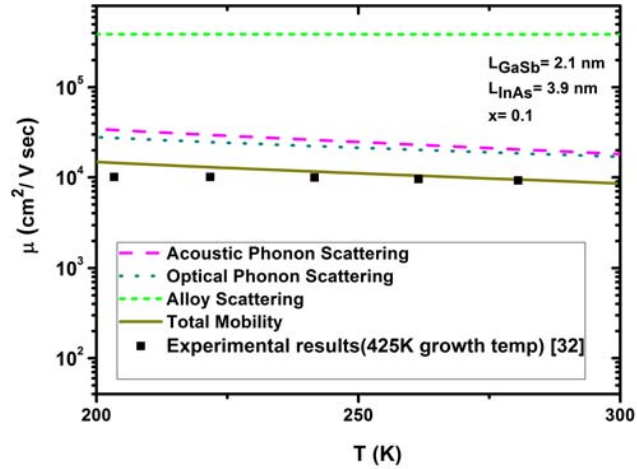

Fig.6. Electron mobility vs. Temperature. (A comparison of our calculation with experimental data taken from Ref [32]).

4. Conclusions

The electron mobility in InAs/GaSb SLs was calculated by solving the corresponding Boltzmann equations using the energy spectra and wave functions obtained from the solution of the K.P Hamiltonian by Finite Difference method. As a result of these calculations, an enhancement in layer thickness leads to semiconductor-semimetal transition, and causes the electron effective mass and penetration of wave function to decrease in magnitude.

As scattering mechanisms were studied in type II SLs, the computations expectedly show that at high temperatures, phonon scattering has a dominant effect on electron mobility. In particular, the reduction in electron mobility is much more due to the powerful effects of optical phonons rather than the acoustic ones. This major effect of optical phonon scattering on limiting the electron mobility becomes more remarkable in thick layer systems; that is because such a systems consist of electrons with a lower energy comparable to the optical phonon energy of the SL.

Also, the Alloy scattering effects on electron mobility in SL interfaces has been considered. We found out that the alloy scattering rate is almost temperature independent. Only a slight variation of alloy scattering versus temperature is revealed which is due to the temperature dependency of the distribution function.

Finally, we have summed over three scattering effects to figure out the variation of total mobility versus temperature. It is realized that thick layer SLs are more suitable for high temperature applications.

A comparison of our calculated results with published experimental data has been made which shows that the theoretical predictions are in good agreement with the experimental outcomes.